\documentclass[letterpaper,11pt]{article}
\pdfoutput=1
\usepackage{graphicx, xspace, subfigure,url}
\usepackage{xcolor}
\usepackage{amssymb,amsmath,amsthm}
\usepackage{multirow}
\usepackage[numbers]{natbib}
\usepackage[left=1in,top=1in,right=1in,bottom=1in]{geometry}
\newcommand{\bi}{\begin{itemize}}
\newcommand{\ei}{\end{itemize}}

\newcommand{\eg}{{\it e.g.,}\xspace}
\newcommand{\ie}{{\it i.e.,}\xspace}
\newcommand\eat[1]{}

\newcommand{\ankitr}[1]{#1}
\newtheorem{thm}{Theorem}[section]

\newtheorem{lemma}[thm]{Lemma}
\newtheorem{claim}[thm]{Claim}

\def\squarebox#1{\hbox to #1{\hfill\vbox to #1{\vfill}}}

\usepackage{framed}
\colorlet{shadecolor}{gray!25}   % you may try 'blue' here
{\normalsize \endMakeFramed}

\title{\vspace{-30pt}On the Resilience of Routing Tables}
\author{Joan Feigenbaum (Yale), Brighten Godfrey (UIUC), Aurojit Panda (UC Berkeley), \\Michael Schapira (Hebrew University), Scott Shenker (UC Berkeley), Ankit Singla (UIUC)}
\date{}

\begin{document}
\eat{
\conferenceinfo{Hotnets '11,} {November 14--15, 2011, Cambridge, MA, USA.}
\CopyrightYear{2011}
\crdata{978-1-4503-1059-8/11/11}
\clubpenalty=10000
\widowpenalty = 10000
}
\maketitle
\thispagestyle{empty}

\begin{abstract}
Many modern network designs incorporate ``failover'' paths into routers' forwarding tables. We initiate the theoretical study of the conditions under which such \emph{resilient routing tables} can guarantee delivery of packets.
\end{abstract}

%!TEX root = paper.tex

\section{Introduction}

The core mission of computer networks is delivering packets from one point to another. To accomplish this, the typical network architecture uses a set of forwarding tables (that dictate the outgoing link at each router for each packet) and a routing algorithm that establishes those forwarding tables, recomputing them as needed in response to link failures or other topology changes. While this approach provides the ability to {\em recover} from an arbitrary set of failures, it does not provide sufficient {\em resiliency} to failures because these routing algorithms take substantial time to reconverge after each link failure.  As a result, for periods of time ranging from 10s of milliseconds to seconds (depending on the network), the network may not be able to deliver packets to certain destinations. In comparison, packet forwarding is several orders of magnitude faster: a 10 Gbps link, for example, sends a $1500$ byte packet in $1.2$ $\mu$sec.

In order to provide higher availability we must design networks that are more resilient to failures. To this end, many modern network designs incorporate various forms of ``backup'' or ``failover'' paths into the forwarding tables that enable a router (or switch), when it detects that one of its attached links is down, to use an alternate outgoing link. We call these {\em resilient routing tables} since they embed failover information into the routing table itself and do not entail changes in packet headers (and so require no change in the low-level packet forwarding hardware). Because these failover decisions are purely local --- based only on the packet's destination, the packet's incoming link, and the set of active incident links --- they occur much more rapidly than the global recovery algorithms used in traditional routing protocols and thus result in many fewer packet losses.

While such resilient routing tables are widely used in practice (\eg ECMP), there has been little theoretical work on their inherent power and limitations. 
 \ankitr{In this paper, we prove that starting with arbitrary loop-free routing tables, we can add forwarding rules to provide resilience against single failures in all scenarios (so long as the network remains topologically connected).} We show, in contrast, that perfect resilience is not achievable in general (\ie there are cases in which no set of routing tables can guarantee packet delivery even when the graph remains connected). We leave open the question of closing the large gap between our positive and negative results. Other interesting open questions include exploring resilient routing tables in the context of specific families of graphs, randomized forwarding rules, and more.

\ankitr{The prior work closest to ours is Failure Insensitive Routing (FIR)~\cite{firProactive}. FIR is also able to guarantee resilience to a single link failure, but is restricted to starting with shortest path routing tables. Our result on resilience to a single failure is more general, allowing the use of arbitrary (loop-free) routing tables in the absence of failure; and adding rules for tolerating one failure. In addition, we also demonstrate the impossibility of perfect resilience. FIR does not discuss a negative result of this nature.}

\ankitr{While there is other significant past research on how to make routing more resilient, these efforts differ from our discussion here in one or more important respects.} For instance, the literature discusses approaches that: (a) use bits in the packet headers to determine when to switch from primary to backup paths (this includes MPLS Fast Reroute)~\cite{idags, kvalbeinMT,mplsfrr}; (b) encode failure information in packet headers to allow nodes to make failure-aware forwarding decisions~\cite{fcp, recycle, hyperbolicgreedy} (work on fault-tolerant compact routing~\cite{ftcompact} also fits in this category); and (c) use graph-specific properties to achieve resilience~\cite{rbgp}. Our own recent work~\cite{ddc-hotnets} provides full resilience (\ie guaranteed packet delivery as long as the network remains connected), but modifies routing tables on the fly.

%!TEX root = paper.tex

\section{Model}

The network is modeled as an undirected graph $G=(V,E)$, in which the vertex set consists of source nodes $\{1, 2, \ldots, n\}$ and a \emph{unique} destination node $d\notin [n]$. Each node $i\in [n]$ has a \emph{forwarding function} $f_i^d:E_i\times 2^{E_i}\rightarrow E_i$, where $E_i$ is the set of node $i$'s incident edges. $f_i^d$ maps incoming edges to outgoing edges as a function of which incident edges are up. We call an $n$-tuple of forwarding functions $f^d=(f^d_1,\ldots,f^d_n)$ a \emph{forwarding pattern}.

Consider the scenario that a set of edges $F\subseteq E$ fails. A \emph{forwarding path} in this scenario is a route in the graph $H^F=(V,E\setminus F)$ such that for every two consecutive edges $e_1,e_2$ on the route which share a mutual node $i$ it holds that $f_i^d(e_1,E_i\setminus F)=e_2$.

%We say that a node $i\in [n]$ is $t$\emph{-connected} to the destination in forwarding pattern $f^d$ if for every failure scenario $F\subseteq E$ such that $|F| \leq t$ there exists a forwarding path from node $i$ to the destination.

Intuitively, our aim is to guarantee that whenever a node is connected to the destination $d$, it also has a forwarding path to the destination. Formally, we say that a forwarding pattern $f$ is $t$\emph{-resilient} if for every failure scenario $F\subseteq E$ such that $|F| \leq t$, (1) if there exists some route from a node $i$ to $d$ in $H^F$ then there also exists a forwarding path from $i$ to $d$ in $H^F$; and (2) all forwarding paths in $H^F$ are loop-free. (Observe that the combination of these two conditions implies, intuitively, that a packet never enters loop en route to the destination or, alternatively, ``gets stuck'' at an intermediate node.)
%\input{related}
%!TEX root = paper.tex

\section{Positive Result}

\subsection{High-Level Overview}

\begin{figure}
  \begin{center}
    \includegraphics[scale=0.5]{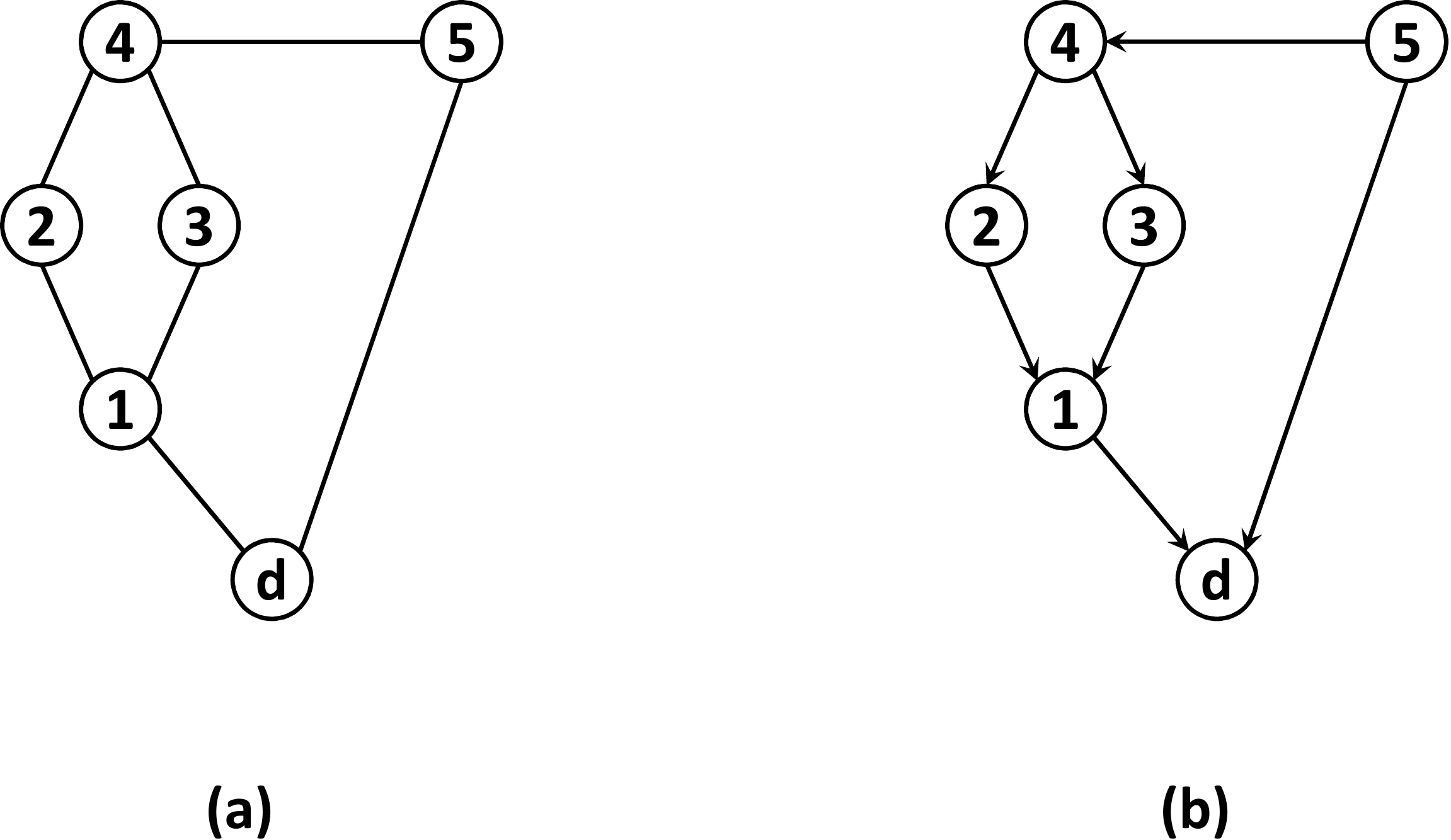}
  \end{center}
  \caption{Illustration of high-level idea}
\end{figure}

We now present our main result, which establishes that for every given network it is possible to efficiently compute a 1-resilient forwarding pattern.

\begin{thm}~\label{thm:t1fd}
For every network there exists a 1-resilient forwarding pattern and, moreover, such a forwarding pattern can be computed in polynomial time.
\end{thm}

We prove Theorem~\ref{thm:t1fd} constructively; we present an algorithm that efficiently computes a 1-resilient forwarding pattern. We now give an intuitive exposition of our algorithm. We first orient the edges in $G$ so as to compute a directed acyclic graph (DAG) $D$ in which each edge in $E$ is utilized. Our results hold regardless of how the DAG $D$ is computed. An example network and corresponding DAG appear in figures 1(a) and 1(b), respectively. The DAG $D$ naturally induces forwarding rules at source nodes; each node's incoming edge in $D$ is mapped to its first active outgoing edge in $D$, given some arbitrary order over the node's outgoing edges (\emph{e.g.}, node 4 in the figure forwards traffic from node 5 to node 2 if the edge to 2 is up, and to node 3 otherwise).
%Note that this does not yet fully specify the forwarding functions (as we have not yet specified to which edges nodes' outgoing edges in $D$ are mapped).

Intuitively, the next step is to identify a ``problematic'' node, that is, a node that is bi-connected to the destination in $G$ but not in the partial forwarding pattern computed thus far, and add forwarding rules so as to ``fix'' this situation. Once this is achieved, another problematic node is identified and fixed, and so on. Observe that nodes 1-4 in the figure are all problematic. Observe also that adding the two following forwarding rules fixes node 4 (\emph{i.e.}, makes node 4 bi-connected to the destination in the forwarding pattern): (a) when both of node 4's outgoing edges in $D$ are down, traffic reaching 4 from node 5 is sent back to 5; and (b) when node 5's direct edge to the destination is up, traffic reaching node 5 from node 4 is sent along this edge. Thus, the algorithm builds the forwarding functions at nodes gradually, as more and more forwarding rules are added to better the resilience of the forwarding pattern.

Implementing the above approach, though, requires care; the order in which problematic nodes are chosen, and the exact manner in which forwarding rules are fixed, are important. Intuitively, our algorithm goes over problematic nodes in the topological order $<_D$ induced by the DAG $D$ (visiting problematic nodes closer to the destination in $D$ first), and when fixing a problematic node $i$, forwarding rules are added until a minimal node in $<_D$ whose entire sub-DAG in $D$ does not traverse $i$ is reached. We prove that this scheme outputs the desired forwarding pattern in a computationally-efficient manner. 

\subsection{Algorithm and Correctness}

\subsubsection{Algorithm}

\begin{enumerate}

\item {\bf Initialize.} $\forall e=(i,j)\in E, \forall T\subseteq E$, set $f^d_j(e,T):=\emptyset$.

\item {\bf Construct DAG.} Construct a DAG $D=(V,E_D)$ (\emph{e.g.}, using BFS/DFS) that is rooted in $d$ and such that $\forall (i,j)\in E$, $(i,j)\in E_D$ or $(j,i)\in E_D$. $D$ induces the following partial order $<_D$ over $V$: $\forall i,j\in V$, $i <_D j$ iff there is a route from $j$ to $i$ in $D$.

\item {\bf Install DAG-based forwarding rules.} $\forall i\in V$, let $E^i_D$ denote the set of $i$'s outgoing edges in $D$. Choose an order over every $E^i_D$ in some arbitrary manner. $\forall j\in V$ such that $e=(j,i)\in E$ and  $\forall T\subseteq E$ such that $T\cap E^i_D\neq \emptyset$ set $f_i(e,T)$ to be the highest element in $E^i_D$ that is not in $T$.

\item {\bf Install additional forwarding rules.} While there exists a node $q$ that is bi-connected to $d$ in $G$ but not in $f^d=(f^d_1,\ldots,f^d_n)$ 
(that is, for which there do not yet exist at least two edge-disjoint forwarding paths to the destination in $f^d$) do:

\begin{enumerate}

    \item Choose $i$ to be a minimal node (under $<_D$) that is bi-connected to $d$ in $G$ but not in $f^d=(f^d_1,\ldots,f^d_n)$.

    \item Choose $j$ to be a minimal node (under $<_D$) such that (1) $i<_D j$ and (2) $\exists x\in V$ such that $(j,x)\in D$ and $i \nleq_D x$.

    \item Choose a simple route $R=(j=v_1,v_2,\ldots,v_k=i)$ from $j$ to $i$ in $D$.
     
    \item Set $c:=k-1$.

    \item While ($c>1$) and ($f_{v_{c}}^d(v_{c+1},v_c)=\emptyset$) do:

    \begin{itemize}
    \item $f_{v_{c}}^d(v_{c+1},v_c):=(v_c,v_{c-1})$
    \item $c:=c-1$
    \end{itemize}

    \item If $c=1$, then $f_{j}^d(v_2,v_1):= (j,x)$.
\end{enumerate}

\end{enumerate}

\subsubsection{Proof of Theorem~\ref{thm:t1fd}}

We now show that the algorithm outputs a forwarding pattern $f^d$ as in the statement of Theorem~\ref{thm:t1fd}. Consider a node $i$ chosen in Step 4b of the algorithm.

\begin{claim}
For every node $i$ that is bi-connected to $d$ in $G$ but not in $f^d$ there exists a node $j$ such that (1) $i<_D j$; and (2) $j$ has a directed edge in $D$ to some node $x$ such that $i \nleq_D x$.
\end{claim}
\begin{proof}
$D$ spans all nodes in $G$ and so there must exist a route $R_1$ from $i$ to $d$ in $D$. $i$ is bi-connected to $d$ in $G$ and so there must also exist another route $R_2$ that is edge-disjoint from $R_1$ and is not in $D$ (otherwise $i$ would be bi-connected to $d$ in $D$). Let $j$ be a node on $R_2$ that has a route $R_3$ to $d$ in $D$ that does not go through $i$. We can now go over the nodes in $R_3$ (from $j$ to $d$) one by one until we reach a node as in the statement of the claim.
\end{proof}

Consider an iteration of Step 4 of the algorithm. Recall that the node $i$ chosen at that iteration is a node that (at that point in time) is bi-connected to $d$ in $G$ but not in $f^d$, and node $j$ is a minimal node such that $i<_D j$ and that has a child $x$ in $D$ for which $i\nleq_D$.

\begin{figure}[!ht]
  \begin{center}
    \includegraphics[scale=0.5]{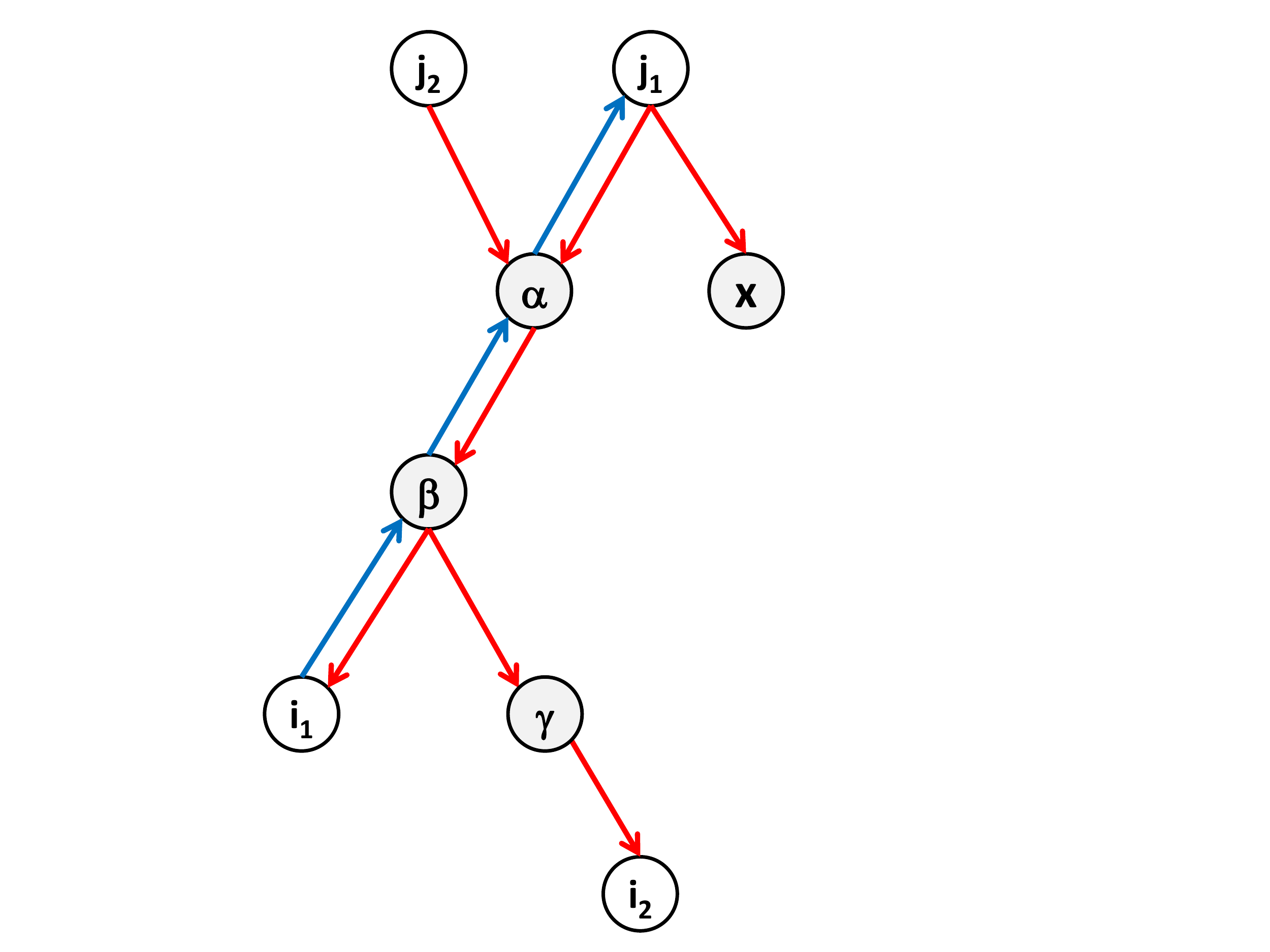}
  \end{center}
  \caption{Illustration of proof idea}\label{fig:positive_proof}
\end{figure}

We now show that following the execution of Step 4 the chosen node $i$ becomes bi-connected to $d$ in $f^d$ and thus ceases to be ``problematic''. We handle two cases.

\begin{itemize}

\item {\bf Case I:} In the execution of Step 4, $c$ is decreased until $c=1$. Observe that in this case $i$ (that already has a route to $d$ in $D$) has (at the end of that iteration) two edge-disjoint forwarding paths to $d$ in $f^d$.

\item  {\bf Case II:} $c$ is decreased until a non-empty ``entry'' in $f^d$ is reached. We now show that in this case, too, $i$ has two edge-disjoint forwarding paths to $d$ in $f^d$ at the end of that iteration.
\end{itemize}

We now handle Case II above. For ease of exposition we illustrate our arguments on the specific (sub)network described in Figure~\ref{fig:positive_proof}. Recall that in Step 2 of the algorithm we construct a DAG $D$. The nodes and the red directed edges in the figure are some subgraph of $D$ (the destination node $d$ does not appear in the figure). Let $i_1$ and $j_1$ be the nodes $i$ and $j$, respectively, chosen at some iteration $q_1$ of Step 2 of the algorithm, and let $R_1=(j_1,\alpha,\beta,i_1)$ be the route $R$ selected at iteration $q$. The blue directed edges in Figure~\ref{fig:positive_proof} represent the changes to the forwarding functions made in the $q_1$'th iteration (along the route $R_1$). Let $i_2$ and $j_2$ be the nodes $i$ and $j$, respectively, selected as some later iteration $q_2>q_1$ of Step 2, and let $R_2=(j_2,\alpha,\beta,\gamma,i_2)$ be the route $R$ selected at iteration $q_2$.

Now, suppose that at the end of iteration $q_1$ node $i_1$ is not only bi-connected to $d$ in $G$ but also in $f^d$. We now show that at the end of the $q_2$'th iteration, $i_2$ too shall be bi-connected to $d$ in both $G$ and $f^d$. Consider the $q_2$'th iteration of Step 2. Observe that at the $q_2$'th iteration $c$ is decreased until it reached the node $\alpha$ as, at that point, a non-empty entry in the forwarding function is reached. Hence, after the $q_2$'th iteration the route $(i_2,\gamma,\beta,\alpha,j_1,x)$ exists in the network. We now show that $i_2\nleq_D x$ and so there exists a route from $i_2$ to $d$ that does not intersect its routes to $d$ in $D$.

By contradiction. Suppose that $i_2\leq_D x$. Recall that $j_1$ was chosen at iteration $q_1$ because it was a minimal node such that $i_1 <_D j_1$ and has a child $x$ in $D$ such that $i\nleq_D x$. Hence, it must be that $i_1 <_D \gamma$ because otherwise $\beta$ would have been chosen instead of $j_1$. Similarly, $i_1 <_D i_2$ because otherwise $\gamma$ would have been chosen instead of $j_1$. This, combined with our assumption that $i_2\leq_D x$ implies that $i_1\leq_D x$ --- a contradiction! The proof of the theorem follows.

\begin{figure}
  \begin{center}
    \includegraphics[scale=0.5]{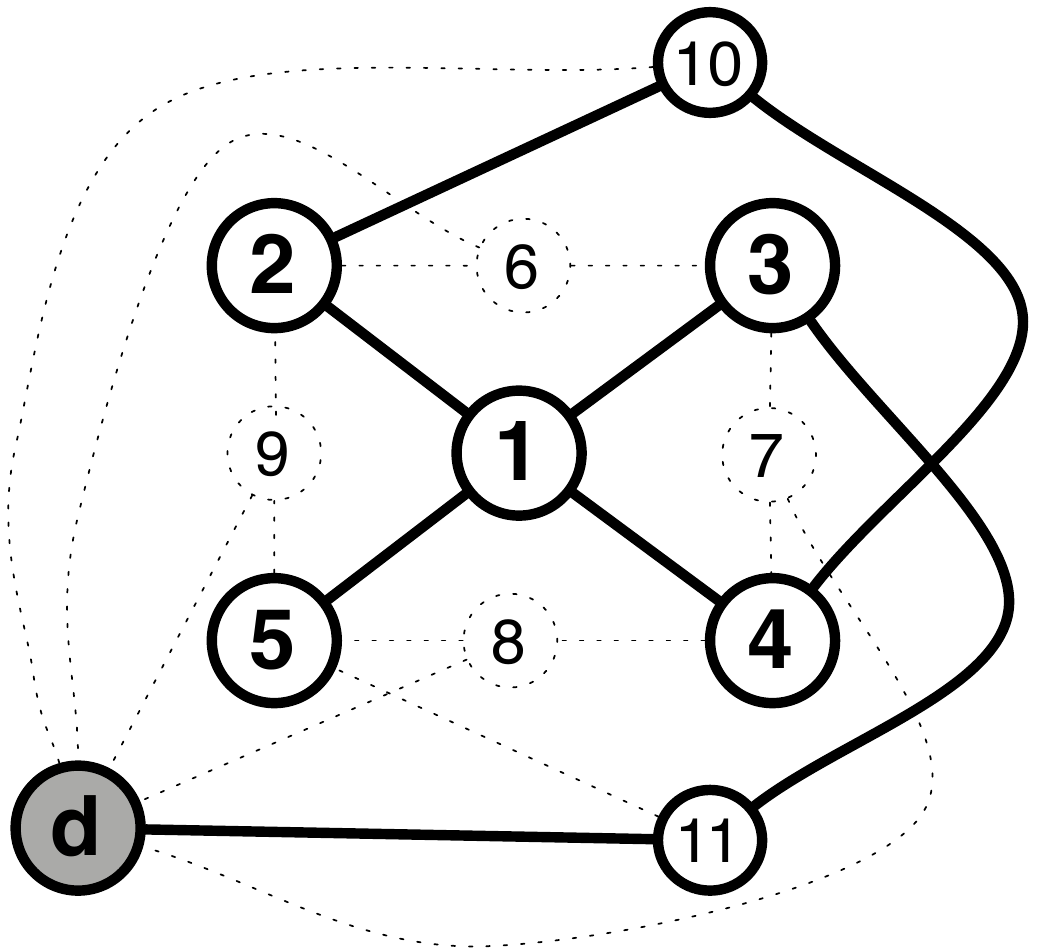}
  \end{center}
  \caption{A failure scenario where perfect resilience is impossible.}\label{fig:negative_proof}
\end{figure}

%!TEX root = paper.tex

\section{Negative Result}

We say that a forwarding pattern $f$ is \emph{perfectly resilient} if it is $\infty$-resilient --- so that regardless of the failure scenario $F \subseteq E$, if there exists some route from a node $i$ to the destination $d$ in $H^F$ then there also exists a forwarding path from $i$ to $d$ in $H^F$.  To prove that forwarding patterns cannot always achieve perfect resilience, we first prove two properties of perfectly resilient forwarding patterns.

\begin{lemma} \label{lemma:first}
For any edge $e_{uv}$, if $v$ has any working path to the destination which does not use the edge $e_{vu}$, then $v$ must not send a packet traveling $u \rightarrow v$ back to $u$. 
\end{lemma}

\begin{proof} Assume the contrary, \textit{i.e.}, there is a perfectly resilient forwarding pattern $f$ with $f_v^d(e_{uv},E_v) = e_{vu}$ and $\exists e_{vw} \in E_v, w \neq u$ such that $w$ has a working path to $d$. Now, consider a scenario where all edges at $u$ other than $e_{uv}$ fail while $v$ is connected to $d$ through $e_{vw}$. A packet from $u$ must be sent to $v$ along $e_{uv}$. Then $f_v^d(e_{uv},E_v) = e_{vu}$ implies $v$ sends the packet back to $u$. $u$ having no other live edges, sends it back to $i$, and we have a forwarding loop, even though there is a route to $d$. This contradicts the claim of $f$ being perfectly resilient.
\end{proof}

\begin{lemma}\label{lemma:second}
A node $i$ in the destination's connected component must route in some cyclic ordering of $E_i \setminus F$, \textit{i.e.}, an ordering of its edges with its neighbors ${v_1, \ldots, v_m}$ such that $\forall j < m: f_{i}(v_j, E_i \setminus F) = v_{j+1}$ and $f_{i}(v_{m}, E_i \setminus F) = v_1$. For example, in figure~\ref{fig:negative_result}, node 1 may route packets from 2 to 3, packets from 3 to 4, from 4 to 5, and from 5 to 2. 
\end{lemma}

\begin{proof}
Let $nbrs(i)$ be the set of neighbors of node $i$. Assume the lemma is false, \textit{i.e.}, there is a perfectly resilient forwarding pattern $f$ such that $f_i$ does not use such a cyclic ordering over $nbrs(i)$. Then $f_i$ must have a smaller cyclic ordering which skips some neighbors $S \subset nbrs(i)$. Consider a scenario where $u \in S$ has a route to $d$, but all edges from nodes in $nbrs(i) \setminus S$ have failed, except those to $i$. The cyclic ordering in $f$ over $nbrs(i) \setminus S$ ensures that packets loop over these nodes: packets starting at any node in $nbrs(i) \setminus S$ are sent to $i$ which forwards them to some other node in the set (per the cyclic ordering). Any such node has no other connectivity except $i$, so the process repeats \emph{ad infinitum}. However, each node in $nbrs(i) \setminus S$ does have a route to $d$ through $u$. This contradicts the claim of $f$ being perfectly resilient.
\end{proof}

\begin{thm}~\label{thm:nfd}
There exists a network for which no perfectly resilient forwarding pattern exists.
\end{thm}

\begin{proof}
Consider the example network in figure (c). We show that after certain failures, no forwarding pattern on the original graph allows each surviving node in the destination's connected component to reach the destination. In figure (c), the surviving links are shown in bold; all other links fail.

By Lemma~\ref{lemma:second} above, node $1$ has to route packets in some cyclic ordering of its neighbors. By the topology's symmetry, we can suppose w.l.o.g. that this ordering is $2, 3, 4, 5, 2,$ \textit{i.e.}, $f^d$ is defined such that $1$ forwards packets from $2$ to $3$, packets from $3$ to $4$, \emph{etc.} Note that a forwarding loop is formed when a packet repeats a directed edge in its path (rather than just a node). To show that this occurs, consider the path taken by packets sent by $5$ after the failures. By Lemma~\ref{lemma:first}, packets sent $1 \to 2$ must not loop back, and so must travel $2 \to 10 \to 4 \to 1$. As a result the packet travels $5 \rightarrow 1 \rightarrow 2 \rightarrow 10 \rightarrow 4 \rightarrow 1 \rightarrow 5 \to 1$ which is a loop since the edge $5 \to 1$ is repeated.
\end{proof}

{\scriptsize
\setlength{\bibsep}{1pt}
\raggedright
\bibliographystyle{abbrvnat}
\bibliography{alinsimple}
}

\end{document}